\documentclass{aastex6}

\usepackage{graphicx}

\begin{document}


\title{A Nonthermal Radio Filament Connected to the Galactic Black Hole?}


\author{Mark R. Morris}
\affil{Dept. of Physics and Astronomy, University of California, Los Angeles, CA 90095-1547, USA}

\author{Jun-Hui Zhao}
\affil{Harvard-Smithsonian Center for Astrophysics, 60 Garden St., Cambridge, MA 02138, USA}

\and

\author{W.M. Goss}
\affil{National Radio Astronomy Observatory, P.O. Box O, Socorro, NM 87801, USA }



\begin{abstract}

Using the Very Large Array, we have investigated a non-thermal radio filament (NTF) recently found very near the Galactic black hole and its radio counterpart, SgrA*.  While this NTF -- the Sgr A West Filament (SgrAWF) -- shares many characteristics with the population of NTFs occupying the central few hundred parsecs of the Galaxy, the SgrAWF has the distinction of having an orientation and sky location that suggest an intimate physical connection to SgrA*.  We present 3.3 and 5.5 cm images constructed using an innovative methodology that yields a very high dynamic range, providing an unprecedentedly clear picture of the SgrAWF.  While the physical association of the SgrAWF with SgrA* is not unambiguous, the images decidedly evoke this interesting possibility.   Assuming that the SgrAWF bears a physical relationship to SgrA*, we examine the potential implications.  One is that SgrA* is a source of relativistic particles constrained to diffuse along ordered local field lines.  The relativistic particles could also be fed into the local field by a collimated outflow from SgrA*, perhaps driven by the Poynting flux accompanying the black hole spin in the presence of a magnetic field threading the event horizon.  Second, we consider the possibility that the SgrAWF is the manifestation of a low-mass-density cosmic string that has become anchored to the black hole.  The simplest form of these hypotheses would predict that the filament be bi-directional, whereas the SgrAWF is only seen on one side of SgrA*, perhaps because of the dynamics of the local medium.
\end{abstract}




\section{Introduction} \label{sec:intro}

The Galactic center HII region, Sgr A West, surrounds the central supermassive black hole, manifested as the radio, infrared, and X-ray source, Sgr A*.   The thermal radio continuum emission from Sgr A West \citep{LoClaussen83, zhao+09} is accompanied by numerous emission lines of hydrogen and helium \citep{RobertsGoss93, scoville+03, paumard+04, qdwang+10}, as well as the 12.8-$\mu$m [Ne II] line \citep*{lacy+91, irons+12}.  The maps and velocity fields of these spectral lines indicate that the ionized gas is organized into more-or-less coherent streams that are responding to the strong gravitational field of the Galactic black hole (GBH), leading to an overall spread of radial velocities exceeding 600 km/s within the central parsec of the Galaxy \citep[e.g.,][]{zhao+09, zhao+10}.  

Sgr A West is surrounded in projection by a supernova remnant, Sgr A East, which exhibits strong, nonthermal radio emission extending over $\sim$3 arcminutes, or $\sim$7 pc \citep*[and references therein]{ZMG16}.
On scales $\ga$ 5 pc, radio observations have also revealed numerous narrow radio filaments, many (but not all) of which have spectral or polarization characteristics indicating that they are illuminated by synchrotron emission.   Many nonthermal radio filaments (NTFs) have been found out to galactocentric distances of $\sim$100 pc \citep[e.g.,][]{YHC04}, and the evidence indicates that they delineate the local magnetic field lines and that the field is highly ordered on large scales \citep[e.g.,][]{morris14}.   

We recently carried out deep 5.5-cm observations of the Sgr A complex with the Karl G. Jansky Very Large Array (VLA) of the NRAO\footnote{The National Radio Astronomy Observatory is a facility of the National Science Foundation, operated under cooperative agreement by Associated Universities, Inc.} that revealed considerable previously unrecognized filamentary structure within 10 pc of Sgr A* \citep{MZG14}.  Most of the filaments are nonthermal, likely consisting of synchrotron emission from relativistic electrons, but Paschen-$\alpha$ images of the region \citep{qdwang+10} show that some of the radio filaments have a counterpart in this recombination line, so they are likely to be thermally emitting features.  Many of the NTFs show filamentary substructure -- a doubling of the filamentary strands, or even in some cases a tightly packed bundle of parallel filaments.  

Until recently, no NTF had yet been found within the central parsec, but \citet{YZ+16a} recently presented radio images of a Òbent filamentÓ located 0.25 - 0.75ÕÕ north of Sgr A*, and one of the suggestions they made for the nature of this feature was that it might be a member of the NTF population.   However, they were not able to demonstrate that the filament is a nonthermal feature because of the difficulty of obtaining its spectral index in this complex region.   \citet{YZ+16a} noted that the bent filament extends toward Sgr A*, but could not be followed closer to Sgr A* than about 5 arcseconds. 

In this paper, we examine the radio filament near Sgr A* with an independent data set.  We subsequently refer to this feature as the Sgr A West Filament, or SgrAWF, and conclude that the SgrAWF is not a thermally emitting structure.  Furthermore, like many NTFs, the SgrAWF is composed of multiple strands.   The relationship of the filament to Sgr A* is examined, and we discuss the possibilities for the origin of this feature.    

\section{Observations} \label{sec:obs}

The continuum observations of the Galactic center were carried out with the VLA at a center frequency of 5.5 GHz (C-band) on 
2012 July 24 and July 27 in the B configuration, under project code 12A-037.   The pointing position was located 35'' east and 10'' south of Sgr A*, and the default correlator setup for broadband continuum observations was used (16 subbands of 64 channels each, with the spectral resolution in each subband being 2 MHz).  The on-source observing time was 5.27 hours.  On 2014 May 17 and May 26, the same setup was used to observe toward the same position with the A configuration under project code 14A-346, with a total of 6.44 hours on source.  

The X-band continuum observations were extracted from the VLA archive; the A-array observations at 9.0 GHz were taken on April 17, 2014, with 5.43 hours pointed at SgrA* (project 14A-232, PI: F. Yusef-Zadeh), again with the same default correlator setup for broadband continuum observations.  The absence of short spacings in the A-array observations led to the presence of artifactual "holes" in the image surrounding the bright features in Sgr A West, which was particularly problematical for this study since portions of the SgrAWF are located in the deepest minimum.   Consequently, we extracted and included a limited amount of B-configuration, X-band data from the archive to reduce this problem: 2.63 hours at 8.6 GHz with just two 128-MHz subbands, each with 64 channels.  These observations were done on 2015  February 20 as part of project 14A-209 (PI: Bower).   While the artifactual holes are ameliorated by the inclusion of B-configuration data, they are still present in both frequency bands, so we have taken care to measure the flux densities reported here with respect to the local background levels.  

In order to obtain the best possible dynamic range in the resulting images, especially in the immediate vicinity of the bright source SgrA*, we found it necessary to do two kinds of corrections, in addition to the standard JVLA calibration as well as the usual phase and amplitude corrections obtained from self-calibration.  First, we corrected for the time variations of Sgr A*, which occur on time scales as short as 5 minutes in the radio, and introduce unacceptable artifacts in the imaging on longer time scales if the variations are not corrected.  The correction procedure for broadband imaging in the presence of such a variable source is discussed in Appendix A of \citet{ZMG16}.  Second, we achieved a substantial reduction in image artifacts by correcting for the antenna-based, spectral-channel-dependent phase errors due to residual delays of a few hundredths of a nanosecond (Zhao et al., in preparation).   Applying this latter correction requires frequent reference to a calibrator source.  For our C-band observations, we used only NRAO 530, which is 16 degrees away from Sgr A*, while for X-band we also used J1744-3116, which is displaced by less than 3 degrees from SgrA*.  Frequent observations of this nearby calibrator have been the key for achieving higher dynamic range in the A-array observations in the X-band.  Finally, the original images created from the first two terms in a Taylor expansion of the intensity spectrum were cleaned using the MS-MFS technique \citep{RauCornwell91}.

The resulting images of the SgrAWF are displayed in Figure 1.   For these images, the "robust" parameter in the CASA data reduction software was set to -1 in C-band and +0.25 in X-band.  The integration time was binned into 6 second intervals for the A-array data at both wavelengths and for the B-array data at 3.3 cm, and to 10 sec. for the B-array data at 5.5 cm.  Spatial smearing due to both finite integration time intervals and spectral channel width is negligible in the regions discussed in this paper.   

Numerous new features emerge from these deep, high-resolution images, most of which will be discussed in separate publications.  We focus here on the NTF located near Sgr A*.   The intensity along the ridge of the filament is shown for both frequencies in Figure 2.  The flux densities have been determined at numerous chosen positions along the filament 
using long rectangular strips oriented perpendicular to the filament in order to determine the local background baseline intensity.  The emission was averaged over the narrow dimension of each strip, the width of which was between 11 and 15 pixels in X-band and 9 to 13 pixels in C-band, depending on the local linear brightness of the filament.  After excluding obvious sources of thermal emission and SgrAWF, a linear fit was made to the flux density along the long axis of each width-averaged strip.  The resulting baseline represented by this fit was subtracted from the width-averaged strip to yield the net local brightness of the SgrAWF.  Along the bright portion of the SgrAWF, the average 5.5 GHz (9 GHz) brightness is 0.48 $\pm$ 0.09 mJy/beam (0.28 $\pm$ 0.06 mJy/beam).  The 5.5 GHz value is much less than that estimated by Yusef-Zadeh et al. (3 - 4 mJy/beam for a similar beam size).  Their 5.5 GHz A-array observations were performed less than 2 months earlier than those presented here, so the difference cannot be ascribed to a real flux density variation because the length of the filament is a few light-years.

Because the filament is unresolved at both frequencies, the orientation of the filament with respect to the elongated synthesized beam affects the measured flux density per beam, so the measurements in mJy/beam were transformed into mJy per arcsecond along the filament by assuming that the filament has a width much smaller than the beam size, and computing the length of the filament segment across the half-maximum contour of the beam.   This procedure allows for a direct comparison of the filament brightness at the two frequencies, in spite of the differing beam sizes. If the filament is partially resolved by the telescope beam, then the linear brightnesses in mJy per arcsecond shown in Figure 2 are lower limits, although the correction factor is close to unity as long as the width of the filament is smaller than the beamwidth. 

\begin{figure*}[ht!] 
\figurenum{1}
\plotone{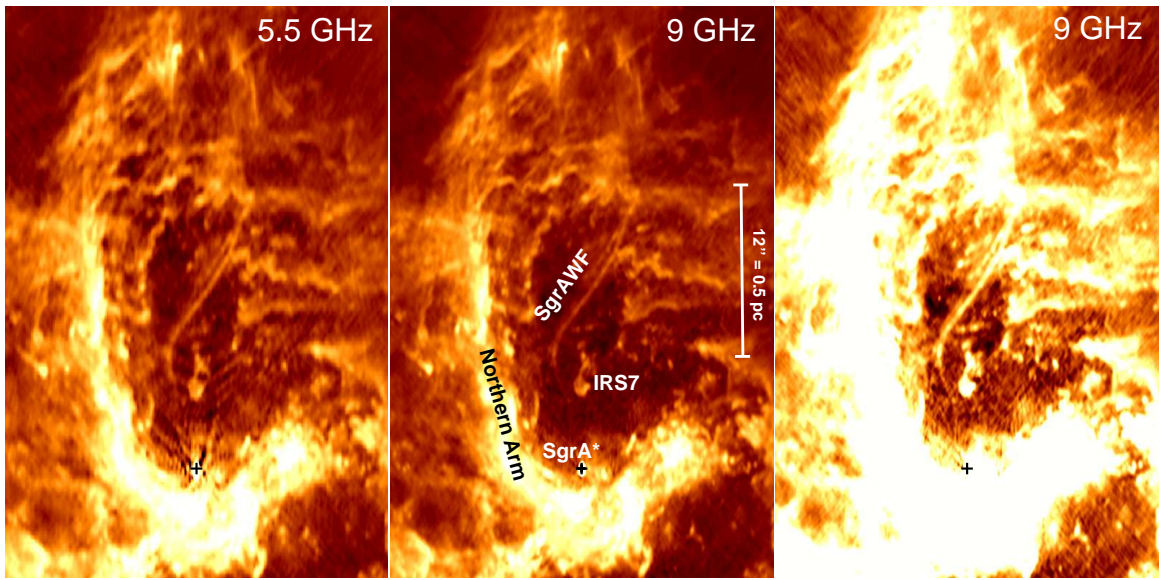} 
\caption{Radiographs of the Sgr A West HII region north of Sgr A*, showing the SgrAW filament. ÊThe black plus sign marks the location of SgrA* in all panels, and all panels have the same scale.  In all images in this paper, equatorial north is at the top.  Ê{\it Center panel}: 9 GHz image, shownÊwith a logarithmic stretch to illustrate the placement of selected features.  Beam size: 0.34'' $\times$ 0.17'', PA = -1.3$^{\circ}$. Ê {\it Left panel}: 5.5 GHz image  stretched to show SgrAWF. ÊThe southernmost portion of the SgrAWF becomes confused by theÊ incompletely  removed  artifacts from the bright emission from Sgr A*, in a region where the dynamic range is $\sim10^4$. ÊBeam size: 0.44'' $\times$ 0.20'', PA = -1.8$^{\circ}$.   {\it Right panel}: 9 GHz image stretched  to show the continuation of the ridge of the SgrAWF to within a few arcseconds of SgrA*, where the dynamic range is $\sim4~\times~10^4$.  Far from Sgr A*, the dynamic range is $3~\times~10^5$ at 5.5 GHz and $3.5~\times~10^5$ at 9 GHz.}
\end{figure*}


\begin{figure*}[ht!]
\figurenum{2}
\plotone{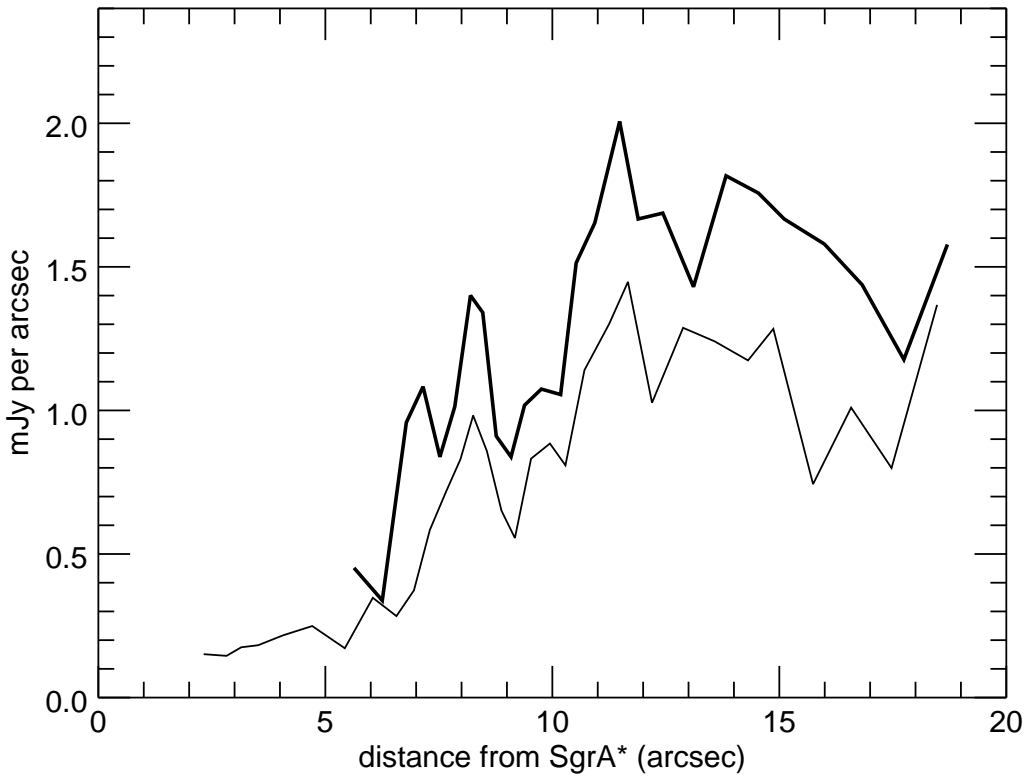}
\caption{Linear brightness of the SgrAWF in X-band (narrow line) and C-band (thick line) in units of mJy per arcsecond, as a function of the direct, linear distance from Sgr A*.  Measurements are shown only where the filament was clearly significant against the local background fluctuations. The conversion from mJy/beam to mJy/arcsecond is described in the text.}
\end{figure*}

\section{Characteristics of the Sgr A West Filament}

Figure 1 shows the C and X-band images of the region containing the Sgr A West filament.  The length of the well-defined portion of the filament that stands out from the complex background is about 0.7 pc.  In its narrow dimension, the filament is unresolved at both frequencies.  As is the case for many NTFs, the brightness of the filament is roughly uniform over much of its length.  However, the brightness declines discontinuously at the southern end of the brighter segment of the filament (at J2000 position 17 45 40.17, -29 00 21.7, which is 6.1 arcsec north of Sgr A*), and the X-band image (Fig 1, right) shows that the filamentary emission continues as a weak ridge that apparently extends to within a few arcseconds of Sgr A*, continuing a linear path that would bring it within $\sim$0.5'' of Sgr A*.  Within a few arcseconds of Sgr A*, the SgrAWF is lost among the remaining residual artifacts near Sgr A*.  Some of these artifacts are evident in the X-band image, and have a brightness that is comparable to the southernmost vestiges of the SgrAWF, but the apparent continuation of the SgrAWF, with its north-south orientation, is not aligned with the predominant direction of the residual linear artifacts, which occur at position angles of $\pm\sim40^{\circ}$.  While improved dynamic range in the immediate vicinity of Sgr A* is clearly required to confirm the association of the SgrAWF with Sgr A*, we assert that the spatial evidence for a direct interaction is already quite compelling.   

The SgrAWF becomes confused with diffuse thermal emission at its northernmost identifiable extent, about 20 arcseconds north of Sgr A*.  It appears to terminate abruptly in an emission knot at J2000 position 17 45 39.75, -29 00 08.7, although this might be coincidental, given the complexity and pervasiveness of the thermal background in that region.  The radio emission knot also shows up in Paschen-$\alpha$ line emission (Figure 3), implying that the radio emission there is predominantly thermal bremsstrahlung.

There are several lines of evidence indicating that SgrAWF is a NTF.  First, a neighboring, parallel strand of emission is present along the northern portion of the brightest strand.  Such filamentary substructure is a common characteristic among NTFs.  Second, the relatively uniform brightness and smooth curvature of the filament contrasts sharply with the emission from the patchy, non-uniformly oriented gas streams of Sgr A West.   Third, and most important, the filament has no counterpart in thermal hydrogen line emission.  Figure 3 shows a Paschen-$\alpha$ image obtained with HST/NICMOS \citep{qdwang+10} placed alongside the radio image.  Paschen-$\alpha$ emission is a bright recombination line in HII regions, but the absence of Pa-$\alpha$ emission toward the filament is a strong indicator that the radio filament is not emitting thermal radiation.  Very large foreground extinction could, in principle, hide a Pa-$\alpha$ counterpart, but the foreground extinction would have to coincide precisely with the filament so that the thermal features that are evident in both the radio and Pa-$\alpha$ images were not also both strongly absorbed.  However, such precise alignment is extremely unlikely.  Fourth, the spectral index, $\alpha$, of SgrAWF is characteristically nonthermal, averaging $-0.75~\pm0.1$ (for flux density $\propto~\nu^{\alpha}$).  This value was obtained using the ratios of the intensities shown in Figure 2, and the resulting ratios are displayed in Figure 4.  The uncertainty of 0.1 in the spectral index is estimated from the dispersion in the measured values of the ratio.  This does not include relative calibration errors between the X and C-band measurements, which in any case contribute a much smaller uncertainty in $\alpha$.  The spectral index of SgrAWF is remarkably constant, indicating that there is no measurable spectral steepening of the emission towards or away from SgrA*.  This is discussed further below.

\begin{figure*}[ht!]
\figurenum{3}
\plotone{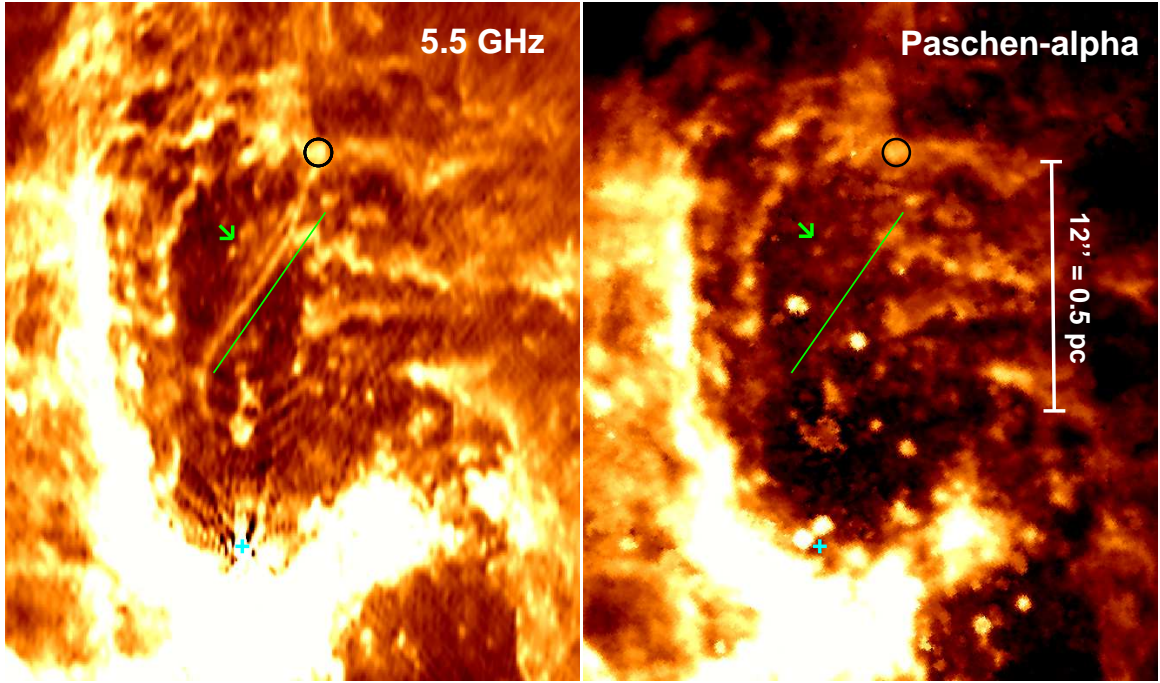}
\caption{Comparison of the 6 cm continuum image (left, same as Fig.\ 1, left) with the Paschen-$\alpha$ image of the same region \citet[][spatial resolution of $\sim$0.2'']{qdwang+10}.  The green line segments (length 8.7'') are drawn for orientation adjacent and parallel to the location of the brightest, relatively straight portion of the radio filament.  The filament has no Pa-$\alpha$ counterpart, in contrast to the thermally-emitting structures that constitute the arms of Sgr A West.  The location of Sgr A* is indicated with a cyan Ò+Ó-sign, and the arrow points to a pair of emission spots that have a radio brightness comparable to that of the radio filament, and which are clearly seen in the Pa-$\alpha$ image, indicating that they are thermal sources.  The black circle in both images surrounds the blob of thermal emission located at the apparent termination point of the SgrAWF.  The point sources appearing in the Pa-$\alpha$ image around Sgr A* are massive, post-main sequence emission-line stars \citep{scoville+03}.}
\end{figure*}



\begin{figure*}[ht!]
\figurenum{4}
\plotone{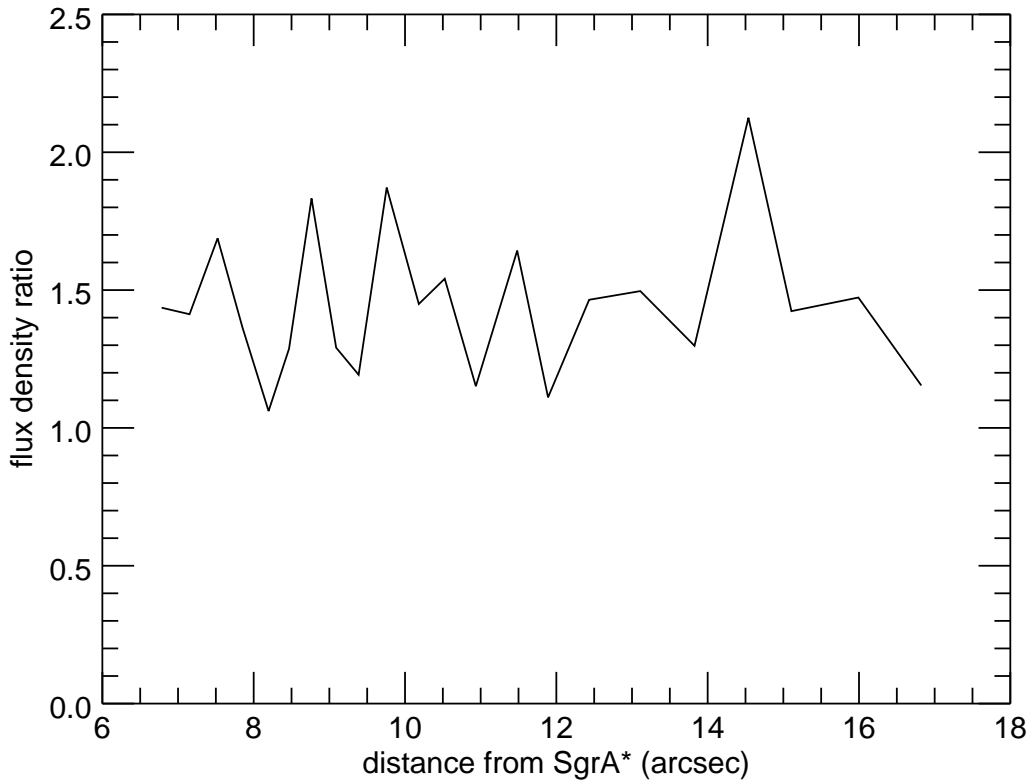}
\caption{Brightness ratio between 5.5 GHz and 9 GHz as a function of distance from Sgr A* along the SgrAWF.  The mean ratio, 1.44, corresponds to a spectral index of -0.75, and we use the dispersion in the measured values of the ratio to estimate an uncertainty in the spectral index of $\pm$0.1. }
\end{figure*}

\section{Discussion} \label{sec:disc}

Perhaps the most important question about the filament is whether it bears a physical relationship with Sgr A*.   The filament could possibly be a fortuitously superimposed member of the population of NTFs that are commonplace in the central 100 pc of the Galaxy \citep{YHC04}.  Indeed, there are at least a dozen NTFs in our 9' $\times$ 9' C-band field of view (Morris, Zhao \& Goss, in preparation).   So a chance superposition cannot be ruled out.  However, there are reasons to suspect that the SgrAWF might be physically associated with Sgr A*: 1) the SgrAWF is the only NTF known to be projected within a parsec of Sgr A*, and therefore the only one projected inside the circumnuclear disk, 2) the end of the brighter portion of the filament at its closest point to Sgr A* points directly at Sgr A*, and 3) while most NTFs have large, slowly varying radii of curvature, the SgrAWF is sharply bent at one location \citep{YZ+16a}.  A small minority of other filaments show such a bending \citep[e.g., the "Snake",][]{Gray95}, which can be attributed to a large localized stress that has deformed the magnetic field that the filament is following.  Furthermore, as synchrotron emitters, all NTFs are necessarily associated with a source of relativistic electrons (or perhaps positrons) somewhere along their length \citep[e.g.,][]{morris96}, and for the SgrAWF, the proximal, accreting supermassive black hole is an obvious candidate for producing such particles.  
These considerations lead us to consider the ways in which the NTF could be physically related to Sgr A*.   

One possibility is therefore that the emitting relativistic particles in the filament are generated by the release of accretion energy in the near vicinity of Sgr A*.  Numerous acceleration processes have been considered to account for the nonthermal radio, sub-millimeter, infrared, and X-ray emission arising from Sgr A* \citep[e.g.,][]{eckart+12, dibi+14, chan+15}; so Sgr A* is clearly a source of such particles.  If a sufficiently strong and ordered magnetic field permeates the region surrounding Sgr A*, that field will confine the escaping particles into a narrow cylinder, shepherding them into a filamentary structure aligned with the field as they diffuse or stream along the magnetic flux tube in which they are trapped.   The synchrotron emission arises as the particles spiral around the field lines in response to the Lorentz force. The guiding of the relativistic particles into a local magnetic flux tube might be facilitated by the Poynting flux resulting from the Blandford-Znajeck mechanism \citep{BZ77}, driven by the spin of the black hole that is threaded by magnetic field lines.  

If Sgr A* is the source of the relativistic particles illuminating the SgrAWF, then the particles should lose energy as they diffuse along the filament.  The cooling time for an electron emitting synchrotron radiation at 9 GHz is $1.4\times10^4/B_{mG}^{3/2}$ years, where $B_{mG}$ is the magnetic field strength in milligauss.  If we assume that particles diffuse along the field lines at the Alfv\'en speed, $v_A=10^{-3}(B_{mG}^2/4\pi\rho)^{1/2}$, where $\rho=m_Hn_b$ is the local mass density, $m_H$ the hydrogen mass, and $n_b$ is the baryon number density, then we can estimate the distance travelled within the cooling timescale -- the diffusion length --  for comparison with the size of the filament.   We adopt the electron density suggested by \citet{Rockefeller+04} for the diffuse, X-ray emitting gas occupying the volume inside the circumnuclear disk, 26 cm$^{-3}$, and find $v_A$ = 430 km s$^{-1}B_{mG}$.  The diffusion length for a relativistic electron emitting at 9 GHz is therefore $\sim6 B_{mG}^{-1/2}$ pc.  Field strength estimates for the central parsec range up to $\gtrsim$8 mG \citep{plante+95, eatough+13}, so the minimum diffusion length is at least a few parsecs, which exceeds the observable length of the filament by a factor of $\sim$ 3.  This factor is consistent with the absence of spectral steepening in our data, but it is small enough to raise the possibility that spectral steepening away from the source of the relativistic particles might eventually be measurable with improved dynamic range.

The northern termination of the SgrAWF in a knot of thermal emission raises the possibility that this knot corresponds to a relatively high-density location in a turbulent plasma at which the relativistic particles in the filament are dumped and deposit their energy as the organizing magnetic field becomes disordered.  This knot of plasma could also be a contributing source of relativistic particles that power the NTF if magnetic reconnection is taking place where the ordered and disordered magnetic fields meet \citep{SM94, uchida+96}.

If the SgrAWF is not merely superimposed on the GC, it can provide crucial information on the magnetic field geometry in the central parsec: the field is ordered, and not turbulent on small scales.  Independent evidence for a relatively ordered magnetic field has resulted from a recent observation of the Faraday rotation of the polarized emission from the nearby magnetar, PSR J1745-2900, which led to the conclusion that a "dynamically relevant" magnetic field exceeding a few milligauss pervades the hot gas within tens of parsecs of the central black hole, and possibly on much smaller scales \citep{eatough+13}.  The bend in the SgrAWF is somewhat unusual among NTFs,  indicating that the field lines have been subject to a substantial local stress where the bend occurs.  A candidate mechanism for bending the SgrAWF would be an oblique shock.  Indeed, a shock front near the location of bend in the SgrAWF has been proposed by  \citep{Rockefeller+05}, who interpret an X-ray emission ridge as the stand-off shock occurring between the blast wave from the nearby supernova remnant, Sgr A East, and the collective winds from the massive stars in the central young star cluster.  The bend in the SgrAWF is located just inside the poorly-defined X-ray ridge, toward Sgr A*, so it could coincide with the leading edge of the advancing shock.

A strong mismatch has been well established between the mass of material that accretes onto Sgr A* through the Bondi radius at about 15,000 AU \citep[10$^{-5}$ - 10$^{-6 }~M_{\sun}~yr^{-1}$;][]{Baganoff+03, qdwang+13} and the mass of material that actually reaches the inner accretion flow at $\sim$1 AU where the radio emission is generated \citep[$\sim10^{-8}~M_{\sun}~yr^{-1}$;][]{Bower+03, marrone+07}.  Consequently, most of the material accreting through the Bondi radius must be ejected in an outflow of some kind \citep[e.g.,][]{sadowski13, yuannarayan14, Yuan+15, RobertsSR+17}.  Both jets and broadly collimated winds have been considered, but the nature of the outflow remains to be observationally determined.  Morphological clues that there is an outflow from the vicinity of Sgr A* abound in deep IR and radio maps \citep{YZM91, muzic+10, ZMG16, YZ+16a}, but direct dynamical evidence has not yet emerged.   Such observationally implied outflows tend not to be highly collimated \citep[although see][]{muzic+10, LMB13, YZ+16a}, and can be ascribed equally well in most cases to the collective stellar winds from the central cluster of young, massive stars.   Consequently, mass outflows attributable to Sgr A* have not yet been unambiguously identified.  Being a nonthermal emitter, the SgrAWF need not have a substantial mass flux, and is therefore not likely to be the primary conduit for the bulk of the outflow from Sgr A* implied by the inflow-outflow mismatch unless the filament is very strongly mass-loaded.  Indeed, with the assumptions that baryons accompany the synchrotron-emitting leptons and that the diffusion rate of all particles is equal to the Alfv\'en speed, the mass outflow rate along the filament is a negligible $3\times10^{-14} B_{mG}^{1/2}$ M$_{\odot}$ yr$^{-1}$.  The curvature and multiplicity of the SgrAWF indicate that it is unlikely to be a relativistic jet.  Although several groups have offered evidence that Sgr A* might be producing a jet \citep[][and references therein]{LMB13}, none of these suggestions has been widely accepted and none of the suggested jets is oriented north-south like the SgrAWF.  In a few cases, the hypothesized jet orientations are perpendicular to the Galactic plane, which differs from the overall orientation of the SgrAWF by $\sim$30$^{\circ}$.

A more exotic interpretation of SgrAWF is a superconducting cosmic string.  As \citet{chudnovsky+86} have argued, light, superconducting, cosmic string loops would migrate toward the center of a Galaxy by frictional deceleration resulting from shock dissipation of their relative velocity as they interact electromagnetically with the magnetized interstellar plasma.  \citet{chudnovsky+86} point out that the large string tension would give rise to string oscillations that lead to extremely high relative velocities of various portions of the string, and of the string segments with respect to the interstellar plasma through which they move.  The high velocity gives rise to a bow shock in the plasma and to a tangential discontinuity surrounding the cosmic string where the pressure of the magnetic field generated by the current in the cosmic string is balanced by the ram pressure of the post-shock plasma.  Near that very-high-pressure discontinuity, the electrons are accelerated to relativistic speeds and can emit synchrotron radiation with sufficient intensity to match that of the Galactic center NTFs.  If the inward migration process for cosmic string loops is carried far enough, a string can presumably migrate close enough to Sgr A* to eventually be captured once a portion of the string passes through the event horizon, leading to a fixed point on the string that alters the oscillatory behavior of the rest of the cosmic string loop.  The problem with the cosmic string hypothesis for any of the known Galactic center NTFs is that the lateral speeds of the oscillating strings are predicted to be a substantial fraction of light speed, yet no such motion has so far been reported.   Future re-imaging of the SgrAWF, even on relatively short time scales, should provide a clear test of the cosmic string hypothesis for this feature.

The hypotheses considered here for a physical connection of the SgrAWF to SgrA* beg the question of why the filament appears on only one side of the black hole.  Indeed, the radio images show no hint of a southern counterpart to the SgrAWF having comparable intensity.  Magnetically confined streams of relativistic particles might be expected to be bisymmetric with respect to the source of the illuminating particles unless the ordered, ambient magnetic field is strongly disturbed in one of the directions.   In the case of the SgrAWF, the magnetic field in the region to the immediate south of SgrA* is likely to be strongly disturbed by the presence of orbiting, ionized gas streams -- the Northern Arm and the Eastern Arm/Bar -- which are both reaching their periapse about 2.7 arcseconds south of Sgr A*, and which could be undergoing a collision at that point \citep{liszt03, zhao+09}, leading to a chaotic velocity field and extreme deformations of any ambient field that would be present there.  Such deformations, combined with the strong thermal emission from the gas streams, would easily mask any nonthermal emission from relativistic particles streaming or diffusing to the south.  A sufficiently high dynamic range image might be able to capture a southern counterpart to the SgrAWF within 1 or 2 arcseconds of Sgr A*, before the dynamical effects of the gas streams intervene.  The absence of a southern counterpart to the SgrAWF might be more problematical for the cosmic string hypothesis (unless the black hole is the end point of the string) although how a string would be manifested in such an environment is unclear.  

\section{Conclusion} \label{sec:cncl}

The Sagittarius A West Filament is a nonthermal radio feature that shares many characteristics with the long-known population of radio filaments that occupy the Galaxy's central molecular zone.  Both the location and orientation of SgrAWF raise the interesting possibility that it is physically linked to the Galaxy's central, supermassive black hole, and that the black hole environment supplies the relativistic particles that illuminate the magnetic filament with their synchrotron radiation. We have also considered the possibility that SgrAWF is a portion of a light, superconducting cosmic string that has been captured by the black hole.  Future observations aimed at achieving higher dynamic range imaging in the immediate vicinity of the bright radio counterpart to the black hole, SgrA*, will be essential for establishing whether SgrAWF is linked to Sgr A*, or is a remarkable but coincidental projection.

\acknowledgments
This work has been supported by NSF grant AST1614782 to UCLA.



\vspace{5mm}
\facilities{VLA}

\software{CASA, MIRIAD, SAOImage DS9}

\end{document}